\documentclass[runningheads]{llncs}

\usepackage{eccv}

\usepackage{eccvabbrv}

\usepackage{graphicx}
\usepackage{booktabs}

\usepackage[accsupp]{axessibility}  

\usepackage{hyperref}

\usepackage{orcidlink}

\begin{document}

\title{OmniForcing: Unleashing Real-time Joint Audio-Visual Generation} 

\titlerunning{OmniForcing: Unleashing Real-time Joint Audio-Visual Generation}


\author{Yaofeng Su\inst{1,2}$^{*}$
\and
Yuming Li\inst{3}$^{*}$ \and Zeyue Xue\inst{1,4} \and Jie Huang\inst{1} \and Siming Fu\inst{1} \and Haoran Li\inst{1} \and Ying Li\inst{3} \and Zezhong Qian\inst{3} \and Haoyang Huang\inst{1} \and Nan Duan\inst{1}}
\authorrunning{Y.~Su et al.}

\institute{
JD Explore Academy \and
Fudan University \and
Peking University \and
The University of Hong Kong 
}

\maketitle
\begingroup
\renewcommand\thefootnote{}\footnotetext{* Equal contribution.}
\endgroup

\begin{abstract}
  Recent joint audio-visual diffusion models achieve remarkable generation quality but suffer from high latency due to their bidirectional attention dependencies, hindering real-time applications. We propose OmniForcing, the first framework to distill an offline, dual-stream bidirectional diffusion model into a high-fidelity streaming autoregressive generator. However, naively applying causal distillation to such dual-stream architectures triggers severe training instability, due to the extreme temporal asymmetry between modalities and the resulting token sparsity. We address the inherent information density gap by introducing an Asymmetric Block-Causal Alignment with a zero-truncation Global Prefix that prevents multi-modal synchronization drift. The gradient explosion caused by extreme audio token sparsity during the causal shift is further resolved through an Audio Sink Token mechanism equipped with an Identity RoPE constraint. Finally, a Joint Self-Forcing Distillation paradigm enables the model to dynamically self-correct cumulative cross-modal errors from exposure bias during long rollouts. Empowered by a modality-independent rolling KV-cache inference scheme, OmniForcing achieves state-of-the-art streaming generation at $\sim$25 FPS on a single GPU, maintaining multi-modal synchronization and visual quality on par with the bidirectional teacher. 
  \noindent\textbf{Project Page:} \href{https://omniforcing.com}{https://omniforcing.com}

\end{abstract}

\section{Introduction}

\begin{figure}[t]
\centering
\includegraphics[width=\textwidth]{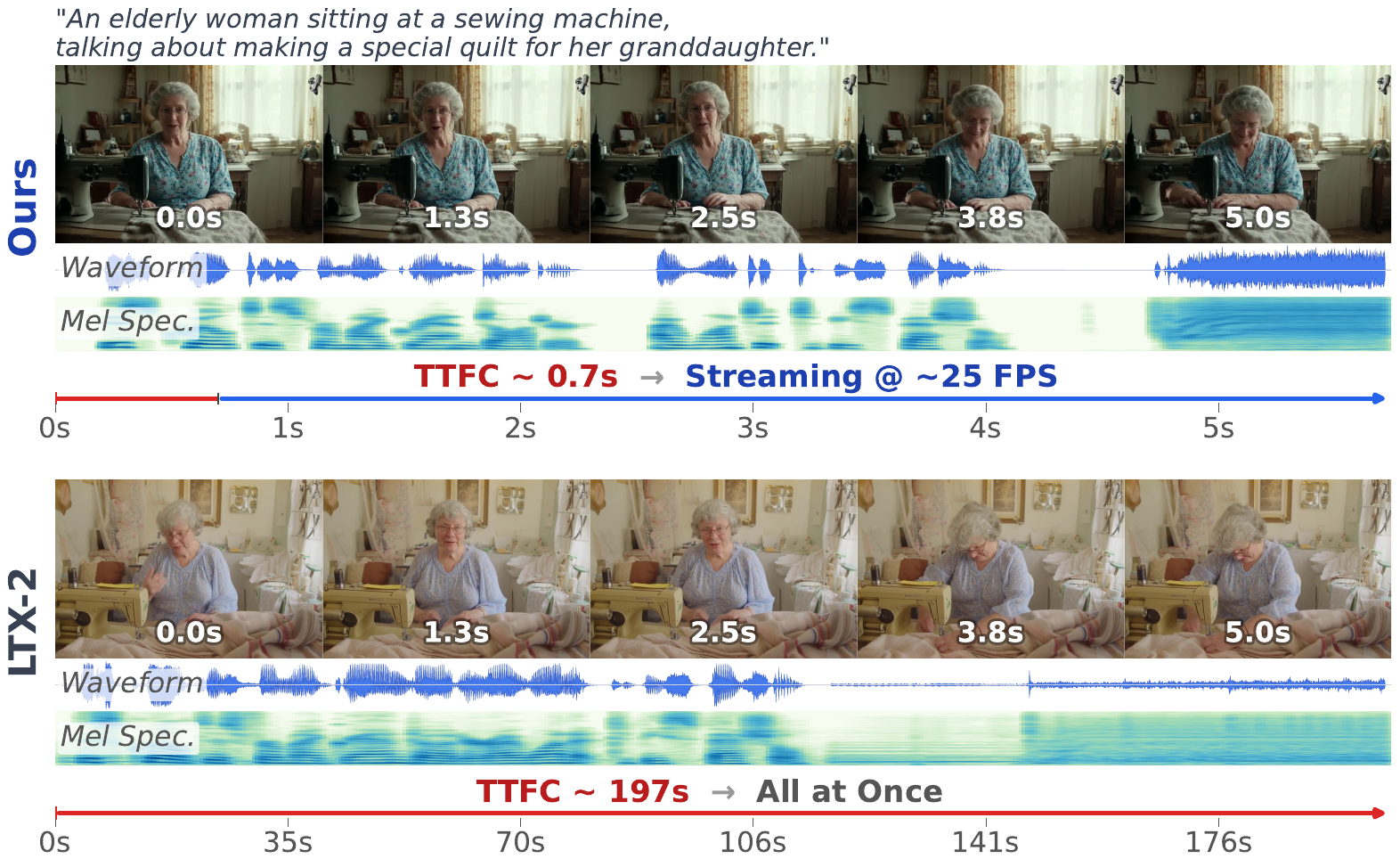}
\caption{OmniForcing breaks the latency barrier for joint audio-visual generation. Top: Our framework achieves real-time streaming at $\sim$25 FPS with an ultra-low Time-To-First-Chunk (TTFC) of $\sim$0.7s. Bottom: The bidirectional teacher (LTX-2) requires $\sim$197s to generate the sequence offline. OmniForcing maintains visual and acoustic fidelity on par with the teacher model.}
\label{fig:teaser}
\end{figure}

The landscape of generative AI has been significantly advanced by Diffusion Transformers (DiTs)~\cite{dit,vaswani2017attention}. Building on this foundation, joint audio-visual models such as LTX-2~\cite{hacohen2026ltx} and Veo 3~\cite{wiedemer2025video} have recently achieved notable progress, leveraging modality-specific VAEs~\cite{kingma2013auto,rombach2022high,liu2023audioldm,hacohen2026ltx,wan2025wan} to map video and audio into continuous latent spaces and jointly model their temporal distribution. However, this capability comes at a substantial computational cost. These models rely heavily on bidirectional full-sequence attention, meaning the entire physical timeline must be processed simultaneously. Consequently, they suffer from a high Time-To-First-Chunk (TTFC) latency~\cite{yin2025slow}, limiting their deployment in interactive, real-time, or streaming applications (see \cref{fig:teaser} for a comparison).

To mitigate this latency bottleneck, previous efforts have diverged into two primary workarounds. The first line employs cascaded pipelines, generating video first and subsequently synthesizing audio~\cite{luo2023diff,cheng2025mmaudio,wang2024frieren,mei2024foleygen} (or vice versa)~\cite{sung2023sound,jeong2023power}. However, this decoupled paradigm severs the joint distribution, limiting generation quality and fundamentally obstructing continuous streaming since the secondary audio modality cannot begin until the primary video has materialized sufficient context. Another line of work adapts video-only diffusion models into causal, autoregressive frameworks (e.g., CausVid~\cite{yin2025slow}, Self-Forcing~\cite{huang2025self}), but these methods remain confined to the visual domain. Directly extending them to dual-stream architectures is non-trivial, as the severe temporal asymmetry between modalities induces a critical information deficit for the sparser modality~\cite{jia2025ditar}, leading to training instability rather than seamless stable alignment.

To address these challenges, we present \textbf{OmniForcing}, the first framework to successfully distill a heavy, bidirectional audio-visual foundation model (\ie, LTX-2~\cite{hacohen2026ltx}) into a high-fidelity streaming autoregressive generator. By dynamically interleaving the generation of audio and video chunks, OmniForcing enables ultra-low latency streaming without sacrificing the holistic multi-modal distribution learned by the bidirectional teacher.

\begin{figure}[t]
\centering
\includegraphics[width=\textwidth]{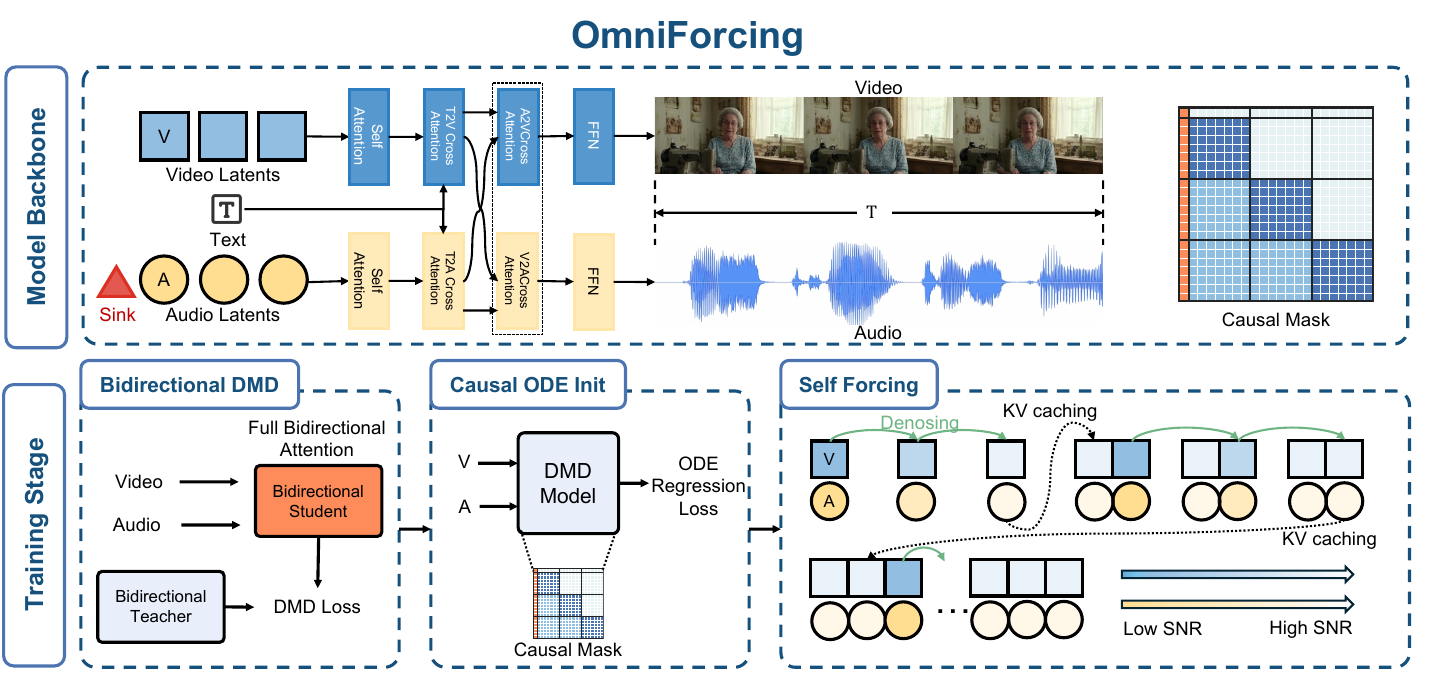}
\caption{The three-stage OmniForcing distillation pipeline. Stage I employs Distribution Matching Distillation (DMD)~\cite{yin2024one,yin2024improved} to adapt the model for few-step, fast denoising. Stage II utilizes causal ODE regression to adapt the network weights to the asymmetric block-causal mask. Stage III implements joint Self-Forcing~\cite{huang2025self} training by autoregressively unrolling the generation process to mitigate exposure bias.}
\label{fig:pipeline}
\end{figure}

Our framework tackles the modality asymmetry through a carefully designed \textit{Asymmetric Block-Causal Alignment}, and the full three-stage distillation pipeline is illustrated in \cref{fig:pipeline}.
Firstly, to establish a persistent cross-modal anchor at the temporal origin, we introduce a zero-truncation \textit{Global Prefix} that aligns the joint sequence at exact one-second boundaries (3 video frames to 25 audio frames) and exploits the native stride characteristics of the VAEs~\cite{hacohen2026ltx,kingma2013auto}.
Then, to stabilize the perilous cross-modal causal shift, we propose an unsupervised \textit{Audio Attention Sink} mechanism, inspired by the attention sink phenomenon in language~\cite{xiao2024efficient} and vision~\cite{darcet2024vision} models. 
Based on this design, we create a position-agnostic global memory buffer by assigning an \textit{Identity RoPE}~\cite{su2024roformer} constraint to these sink tokens, which mitigates the gradient explosions inherent in sparse causal audio attention.
Furthermore, to combat the exposure bias amplified by cross-modal error accumulation during long rollouts, we employ a joint Self-Forcing~\cite{huang2025self} distillation strategy that enables the model to dynamically self-correct. Finally, we exploit the intra-layer decoupling between the 14B video and 5B audio streams~\cite{hacohen2026ltx} by introducing a \textit{Modality-Independent Rolling KV-Cache} that reduces per-step context complexity to $\mathcal{O}(L)$ and enables concurrent execution of the two modality streams on a single GPU.

In summary, our main contributions are:
\begin{itemize}
    \item We propose \textbf{OmniForcing}, a unified autoregressive framework that transforms offline, bidirectional joint audio-visual models into real-time streaming engines while preserving exact multi-modal temporal synchronization.
    \item We introduce a natural \textbf{Asymmetric Block-Causal Alignment} and the \textbf{Audio Sink Token} mechanism with Identity RoPE, providing a robust, position-agnostic solution to the Softmax collapse caused by multi-modal token density mismatch.
    \item We introduce a \textbf{Modality-Independent Rolling KV-Cache} with asymmetric parallel inference and a \textbf{Joint Self-Forcing Distillation} paradigm, which together mitigate exposure bias and reduce per-step context complexity to $\mathcal{O}(L)$, achieving state-of-the-art streaming generation at $\sim$25 FPS on a single GPU.
\end{itemize}

\section{Related Work}

\noindent\textbf{Joint Audio-Visual and Video Foundation Models.}
The landscape of generative AI has been significantly advanced by large-scale Diffusion Transformers (DiTs)~\cite{dit,vaswani2017attention}. For visual generation, foundation models such as Sora~\cite{liu2024sorareviewbackgroundtechnology}, Wan~2.1~\cite{wan2025wan}, HunyuanVideo~\cite{wu2025hunyuanvideo} and Kling~\cite{team2025kling} have demonstrated high visual fidelity, physical realism, and adherence to complex text prompts. Building upon these unimodal visual successes, the field has recently achieved a notable shift toward unified multimodal generation. Joint audio-visual foundation models like LTX-2~\cite{hacohen2026ltx} and Veo~3~\cite{wiedemer2025video} have emerged as state-of-the-art systems capable of generating highly synchronized, high-fidelity audio and video in a single pass. Notably, LTX-2 employs an asymmetric dual-stream architecture (a 14B video stream and a 5B audio stream) coupled through bidirectional cross-attention to deeply model the joint distribution of both modalities. 

While these foundation models deliver remarkable semantic alignment and generation quality, they exhibit a critical limitation regarding deployment: they rely exclusively on \textit{bidirectional full-sequence attention}. Because the generation of a single frame requires the model to simultaneously attend to the entire physical timeline, the computational complexity scales quadratically with sequence length. This results in a massive Time-To-First-Chunk (TTFC) latency, rendering these models ill-suited for supporting real-time, interactive, or streaming applications.

\noindent\textbf{Audio-Visual Synthesis and Alignment.}
Prior to the emergence of joint foundation models, multi-modal generation heavily relied on cascaded or decoupled pipelines. These methods typically generate video first and subsequently synthesize the matching audio track using Foley sound generators (\eg, FoleyGen~\cite{mei2024foleygen}, Diff-Foley~\cite{luo2023diff}, FoleyCrafter~\cite{zhang2026foleycrafter}, MMAudio~\cite{cheng2025mmaudio}), or build upon standalone audio foundation models like AudioLDM~\cite{liu2023audioldm} and AudioGen~\cite{kreuk2023audiogen}. Conversely, other works explore generating video driven by audio signals (A2V)~\cite{sung2023sound,jeong2023power}. 

While computationally tractable, this decoupled paradigm inherently severs the joint temporal distribution. It struggles with fine-grained cross-modal synchronization and complex temporal reasoning, such as visual actions dynamically reacting to sudden acoustic events. Furthermore, this sequential video-to-audio (V2A) or audio-to-video (A2V) dependency fundamentally obstructs real-time streaming, a limitation OmniForcing avoids by streaming both modalities synchronously.


\noindent\textbf{Diffusion Distillation for Efficiency.}
To break the latency barrier, various distillation methods compress multi-step diffusion sampling into one or a few evaluations. Distribution Matching Distillation (DMD)~\cite{yin2024one,yin2024improved} minimizes an approximate KL divergence between student and teacher; Consistency Models~\cite{song2023consistency,luo2023latent} enforce self-consistency along ODE trajectories; and Adversarial Diffusion Distillation~\cite{sauer2024adversarial} leverages discriminator-based losses. These methodologies provide the computational foundation for real-time generation.

\noindent\textbf{Autoregressive \& Streaming Diffusion Models.}
Building upon efficient few-step distillation, recent pioneering works have transformed offline diffusion models into streaming architectures. Early explorations like StreamingT2V~\cite{henschel2025streamingt2v} and Pyramid Flow~\cite{pyramid} introduced frame-wise and pyramid-based autoregressive diffusion, yet remained constrained by multi-step sampling overhead. CausVid~\cite{yin2025slow} first established the core paradigm for streaming diffusion by distilling a bidirectional video teacher into a causal student using an asymmetric DMD pipeline, achieving $\sim$9.4 FPS streaming generation. Following this, Self-Forcing~\cite{huang2025self} identified and solved the critical \textit{exposure bias} problem in autoregressive video generation by forcing the model to unroll its own KV-cache predictions during training. This foundation has rapidly catalyzed a family of ``forcing'' variants tailored for diverse autoregressive bottlenecks, including Causal-Forcing~\cite{zhu2026causal} for stricter causal consistency and Rolling-Forcing~\cite{liu2026rolling} for minute-level long-context generation. 

However, while these works represent a significant step forward for streaming generation, they operate exclusively on unimodal (video-only) architectures. Achieving real-time streaming for \textit{joint audio-visual} generation remains an open, highly compelling, and unresolved problem. Furthermore, naively porting these causal distillation paradigms to a dual-stream multimodal architecture leads to severe training instability. Due to the severe frequency asymmetry between audio and video (e.g., 25 FPS vs. 3 FPS), imposing a causal mask creates extreme token sparsity and severe conditional distribution shifts, triggering Softmax collapse and gradient explosions. Therefore, formulating a stable, architecture-aware distillation pipeline tailored specifically for joint audio-visual streaming is of paramount importance. Our proposed OmniForcing naturally addresses this exact gap.

\section{OmniForcing}
\subsection{Problem Formulation and The OmniForcing Pipeline}

Given a text prompt $c$, our goal is real-time, streaming joint generation of a temporally aligned video $\mathbf{V}$ and audio $\mathbf{A}$, mapped into independent latent spaces via modality-specific VAEs~\cite{kingma2013auto,rombach2022high,liu2023audioldm,hacohen2026ltx}. Following the distillation paradigm established by CausVid~\cite{yin2025slow} and Self-Forcing~\cite{huang2025self}, we restructure a pretrained bidirectional dual-stream transformer (LTX-2~\cite{hacohen2026ltx}) into a \textbf{block-causal autoregressive} framework, factorizing the joint distribution over $K{+}1$ synchronized blocks $\{\mathcal{B}_0, \ldots, \mathcal{B}_K\}$, where $K$ is the total number of physical seconds generated:
\begin{equation}
    p(\mathbf{V}, \mathbf{A} \mid c) = p(\mathcal{B}_0 \mid c) \prod_{k=1}^{K} p(\mathcal{B}_k \mid \mathcal{B}_{<k}, c).
\end{equation}
This retrofit faces three core challenges: (i) the extreme frequency asymmetry (3 FPS video vs.\ 25 FPS audio) hinders conventional causal masking; (ii) restricting the global bidirectional receptive field to sparse causal history triggers Softmax collapse and gradient explosions, an instability that is disproportionately severe for the audio stream, as diminishing the bidirectional context to extremely few tokens fundamentally undermines the modeling of continuous tokens~\cite{jia2025ditar}; (iii) exposure bias during long rollouts~\cite{huang2025self} is amplified into cross-modal desynchronization. OmniForcing addresses all three through an Asymmetric Block-Causal Masking design coupled with a three-stage distillation pipeline, transferring the teacher's high-fidelity joint distribution to an ultra-fast causal engine.

\subsection{Asymmetric Block-Causal Alignment and Mask Design}
\label{sec:mask_design}

To achieve the block-level autoregressive factorization of the joint distribution, we must first bridge the inherent information density gap between the two modalities. In the physical world, video and audio exhibit starkly different spatiotemporal characteristics: video typically presents strong spatial redundancy and a low temporal evolution frequency, whereas audio is a dense high-frequency one-dimensional temporal signal. Consequently, when compressing into the latent space, the VAEs in multimodal generative models inevitably employ highly asymmetric temporal downsampling rates~\cite{hacohen2026ltx}. Specifically, within our LTX-2~\cite{hacohen2026ltx} backbone, the video VAE outputs $f_v = 3$ latent frames per second, while the audio VAE outputs $f_a = 25$ frames per second.

Faced with this 25:3 non-integer frequency ratio, a strict frame-by-frame causal mask would lead to destructive feature truncation and temporal misalignment. To naturally resolve this conflict, we establish a physical-time-based \textbf{Macro-block Alignment}: given a one-second temporal window $\Delta T = 1\,\text{s}$, it perfectly encapsulates $\Delta N_v = f_v \times \Delta T = 3$ video latents and $\Delta N_a = f_a \times \Delta T = 25$ audio latents, without any fractional remainders.

\noindent\textbf{The Global Prefix Token and the Mathematical Fit of VAE Strides.} Notably, this block-level partitioning aligns closely with the causal convolutional stride characteristics of the underlying VAEs. In temporal compression, standard causal VAE architectures~\cite{kingma2013auto, hacohen2026ltx,wan2025wan} typically apply an asymmetric treatment with a stride of 1 for the absolute first frame, while utilizing full-receptive-field strides for subsequent frames (e.g., a stride of 8 for video and 4 for audio). Therefore, the length $N$ of the entire latent sequence strictly follows the derived formulas:
\begin{equation}
    N_v = 1 + K \cdot f_v, \quad N_a = 1 + K \cdot f_a,
\end{equation}
where $K$ is the total number of physical seconds generated (i.e., the total number of blocks). The constant term $1$ in the formula corresponds to the initial latents $\mathbf{V}_0$ and $\mathbf{A}_0$ at $t \approx 0$ seconds.

Based on this derivation, the initial components $\mathbf{V}_0$ and $\mathbf{A}_0$ are naturally anchored at the origin in physical time, making it inherently impossible to squeeze them into the subsequent 1-second standard blocks $\mathcal{B}_k$. We transform this architectural inevitability into a design advantage by explicitly merging them into a \textbf{Global Prefix} block ($\mathcal{B}_0$). Within $\mathcal{B}_0$, the attention mechanism is unconditionally bidirectional. $\mathcal{B}_0$ functions similarly to a system prompt in large language models; it is immune to standard causal decay and remains globally visible to all future tokens. This not only ensures an elegant zero-truncation perfect alignment across the entire sequence but also provides a robust cross-modal semantic anchor for autoregressive long-sequence generation.

\begin{figure}[!t]
\centering
\includegraphics[width=\linewidth]{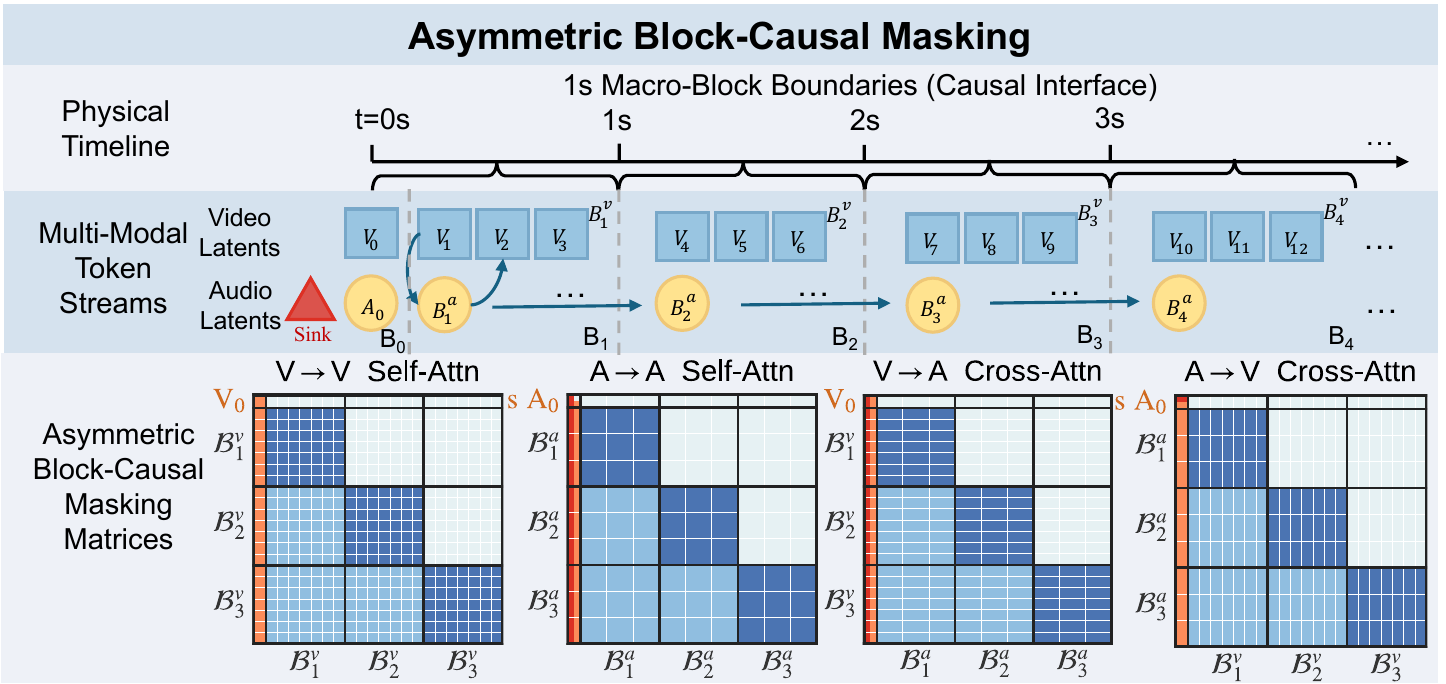}
\caption{Asymmetric Block-Causal Masking. The vertical axis denotes query tokens and the horizontal axis denotes key tokens. Modalities are synchronized via 1s macro-blocks. Each audio block ($B^a$) contains 25 latent frames (one token each), whereas each video block ($B^v$) contains 3 latent frames patchified into $3 \times 384$ tokens. Unmasked tokens include the Global Prefix (orange, $V_0/A_0$) and Audio Sink tokens (red, $s$). Blue regions denote allowed attention (bidirectional intra-block, strictly causal inter-block), while white regions mask future keys to prevent information leakage.}
\label{fig:causal_mask}
\end{figure}
\noindent\textbf{Natural Derivation of Four-Way Asymmetric Causal Masks.} 
Having established the block-level alignment, we formalize the four-way attention masking strategy. Let $\tau(q)$ denote the physical block index of a given token $q$. For the video stream, each latent frame is spatially patchified into $H_v W_v$ tokens, where $H_v, W_v$ are the spatial dimensions after VAE compression and patch embedding. Each standard video block thus contains $|\mathcal{B}^v| = 3 \times H_v W_v$ tokens. For the audio stream, the mel-frequency bins are flattened into the channel dimension by the audio patchifier, so each latent frame maps to a single token; each standard audio block therefore contains $|\mathcal{B}^a| = 25$ tokens. Accounting for the Global Prefix $\mathcal{B}_0$ (which holds $1 \times H_v W_v$ video tokens and $1$ audio token), the block assignment for tokens beyond the prefix is:
\begin{equation}
    \tau_v(q) = 1 + \left\lfloor \frac{q - H_v W_v}{3 \times H_v W_v} \right\rfloor, \quad
    \tau_a(q) = 1 + \left\lfloor \frac{q - 1}{25} \right\rfloor,
\end{equation}
where tokens with indices below the prefix boundary ($q < H_v W_v$ for video, $q < 1$ for audio) belong to $\mathcal{B}_0$.

To enforce strict causality without future information leakage while allowing intra-block bidirectional flow, we define the binary attention mask $\mathbf{M} \in \{0, 1\}$ for query token $q$ and key token $k$ across all four attention pathways:
\begin{enumerate}
    \item \textit{Intra-modal Self-Attention:} 
    \begin{equation}
        \mathbf{M}^{V \to V}_{q, k} = \mathbb{I}(\tau_v(k) \le \tau_v(q)), \quad \mathbf{M}^{A \to A}_{q, k} = \mathbb{I}(\tau_a(k) \le \tau_a(q)),
    \end{equation}
    \item \textit{Cross-modal Attention:} 
    \begin{equation}
        \mathbf{M}^{V \to A}_{q, k} = \mathbb{I}(\tau_a(k) \le \tau_v(q)), \quad \mathbf{M}^{A \to V}_{q, k} = \mathbb{I}(\tau_v(k) \le \tau_a(q)),
    \end{equation}
\end{enumerate}
where $\mathbb{I}(\cdot)$ is the indicator function. Because the Global Prefix tokens $\mathbf{V}_0$ and $\mathbf{A}_0$ are assigned to block $\tau=0$, they inherently satisfy $\tau(k) \le \tau(q)$ for all subsequent queries $q \ge 0$, mathematically guaranteeing their status as a globally visible, unmasked semantic anchor. This formulation ensures that despite the severe token density mismatch, the temporal receptive fields of both modalities expand synchronously at the physical block boundaries. The resulting four-way mask is visualized in \cref{fig:causal_mask}.

\subsection{Bridging the Gap: Causal Regression and Architectural Stabilizers}

Inspired by recent video-only autoregressive distillation works (e.g., CausVid~\cite{yin2025slow}), we design the first two stages of the pipeline. The core idea of these two stages is to smoothly decouple the few-step denoising capability from the causal generation paradigm and inject them into the model sequentially, paving the way for subsequent real-time joint inference.

\noindent\textbf{Stage I: Bidirectional DMD.} We first employ Distribution Matching Distillation (DMD)~\cite{yin2024improved,yin2024one} to distill the original pretrained model into a bidirectional student model requiring very few steps. The loss is a weighted sum of the video and audio score-matching objectives: $\mathcal{L}_{\text{Bi-DMD}} = \lambda_v \mathcal{L}_{\text{DMD}}^v + \lambda_a \mathcal{L}_{\text{DMD}}^a$.
While preserving the original global attention receptive field, this stage endows the model with strong few-step denoising capabilities, thereby providing a high-quality, easily regressible teacher trajectory for the subsequent causal architectural migration.

\noindent\textbf{Stage II: Causal ODE Regression.} Next, we equip the model with the block-causal masks defined in \cref{sec:mask_design}. To adapt the weights to the causal masking without the complexity of full generation, we regress the ODE trajectories of the Stage I teacher $v_{\phi}$. Let $\mathbf{x}_t = [\mathbf{V}_t, \mathbf{A}_t]$ denote the joint noisy latent at flow-matching time $t$, and superscripts $v, a$ denote the video and audio velocity predictions, respectively:
\begin{equation}
    \mathcal{L}_{\text{ODE}} = \mathbb{E}_{t, \mathbf{x}_t} \left[ \lambda_v \| v_\theta^v(\mathbf{x}_t, c) - v_\phi^v(\mathbf{x}_t, c) \|_2^2 + \lambda_a \| v_\theta^a(\mathbf{x}_t, c) - v_\phi^a(\mathbf{x}_t, c) \|_2^2 \right].
\end{equation}
This stage aims to correct the causal maladaptation of the model weights, teaching the model to perform effective denoising predictions by observing only the causal history.

\noindent\textbf{Conditional Distribution Shift and Gradient Explosion Crisis.} Crucially, directly applying causal masks to the dual-stream model leads to a catastrophic collapse of Stage II training. The root cause lies in the severe \textit{conditional distribution shift} when transforming bidirectional pretrained knowledge into the causal domain. The conditional distribution abruptly shifts from a globally-informed posterior to a truncated causal one:
\begin{equation}
    \underbrace{p(\mathbf{x}_i \mid \mathbf{x}_{1:N}, c)}_{\text{Bidirectional (pretrained)}} \;\longrightarrow\; \underbrace{p(\mathbf{x}_i \mid \mathbf{x}_{1:i}, c)}_{\text{Causal (target)}}.
\end{equation}
This information deficit is asymmetric across modalities. Because video utilizes spatiotemporal patch partitioning, a single physical block still contains hundreds of tokens (e.g., $3 \times 384$ for our configuration), possessing a relatively abundant local context. Audio, however, usually has far fewer tokens per block; in our setting it contains a mere $\Delta N_a = 25$ tokens per block.

This extreme token sparsity destabilizes the attention mechanism. For audio tokens in early blocks, the visible history length is exceedingly small---the first token of a new block can only attend to itself and a few preceding tokens. With such a minuscule normalization denominator, the Softmax distribution degenerates into a near-one-hot vector with entropy approaching zero. In this saturated regime, minor logit perturbations are sharply amplified through the exponential nonlinearity, causing gradient variance to surge explosively ($\|\nabla\mathcal{L}\| \to \infty$) and producing NaN losses under fp16/bf16 precision.

\noindent\textbf{Architectural Stabilizer: Audio Sink Tokens with Identity RoPE.} To address this instability at its root, we introduce an effective architectural stabilizer, inspired by the attention sink phenomenon observed in autoregressive language modeling~\cite{xiao2024efficient} and vision models~\cite{darcet2024vision}. We prepend $S$ learnable Sink Tokens to the front of the audio sequence and permanently anchor them within the global prefix ($\mathcal{B}_0$).

In a physical sense, they act as a soft global memory buffer; mathematically, they forcefully expand the attention denominator for early audio tokens from $i$ to $i+S$. This operation successfully breaks the extremely sharp Softmax collapse and restores attention entropy, thereby acting as a contextual buffer to absorb anomalous logit perturbations and quelling the gradient storm at its source.

Furthermore, to prevent standard Rotary Position Embeddings (RoPE)~\cite{su2024roformer} from imparting spurious physical temporal biases to these abstract tokens, we specifically enforce an \textit{Identity RoPE} constraint on them:
\begin{equation}
    \cos(\theta_{\text{sink}}) = \mathbf{1}, \quad \sin(\theta_{\text{sink}}) = \mathbf{0}.
\end{equation}
Under this constraint, the standard rotary transformation analytically degenerates into an identity mapping ($\text{RoPE}(\mathbf{x}) = \mathbf{x}$). This shields the Sink Tokens from any positional rotational interference, making them position-agnostic semantic anchors that help stabilize the model across the causal transition.

\subsection{Joint Self-Forcing Distillation and Asymmetric Parallel Inference}

Although Stage II successfully injects causal generation capabilities, autoregressive models are inherently subject to exposure bias during long-sequence rollouts: because the model must condition on its own imperfect past outputs (rather than ground-truth data) during inference, prediction errors rapidly accumulate.

\noindent\textbf{Stage III: Causal DMD with Joint Self-Forcing.} 
To mitigate exposure bias, we introduce a joint audio-visual Self-Forcing~\cite{huang2025self} paradigm. Rather than conditioning on ground-truth history, the model unrolls the sequence autoregressively during training. Let $G_\theta$ denote the causal student generator and $R_\phi$ the frozen bidirectional teacher. Let $\hat{\mathcal{B}}_k$ be the block generated by $G_\theta$; its KV embeddings are computed without noise and appended to a rolling cache. The joint Self-Forcing DMD loss evaluates the generated trajectory against $R_\phi$, where $\mathbf{z}_k$ is the sampled noise for block $k$:
\begin{equation}
    \mathcal{L}_{\text{SF}} = \sum_{k=1}^K \mathbb{E}_{\hat{\mathcal{B}}_{<k}} \left[ \nabla_\theta \text{KL} \big( G_\theta(\mathbf{z}_k \mid \text{KV}_{<k}, c) \;\parallel\; R_\phi(\mathbf{z}_k \mid c) \big) \right].
\end{equation}
This coupled unrolling mechanism forces the video and audio streams to dynamically adapt to each other's prediction drifts, ensuring strict cross-modal synchrony.

\noindent\textbf{Asymmetric Compute Allocation and Parallel Inference.} Ultimately, OmniForcing achieves true real-time inference by combining the aforementioned autoregressive algorithm with a flexible parallel strategy. In our architecture, the 14B video branch and the 5B audio branch exhibit extreme compute asymmetry. Thanks to our decoupled design in the attention pathways, there is absolutely no data dependency between video self-attention and audio self-attention.

Within each transformer layer, the audio and video streams are computed through independent FFN sub-layers and only synchronize briefly at the cross-modal attention boundaries (A2V and V2A). This intra-layer decoupling, combined with a Modality-Independent Rolling KV Cache that reduces per-step context complexity to $\mathcal{O}(L)$---where $L$ denotes the total number of latent frames within the cache window---enables real-time synchronized audio-visual generation at $\sim$25 FPS on a single GPU. Furthermore, since the two modality-specific FFN sub-layers within each layer share no data dependency, the architecture naturally lends itself to asymmetric tensor parallelism across devices—assigning more compute to the heavier video stream—providing a practical scaling path toward higher resolutions and longer sequences in multi-GPU deployments.

\section{Experiments}

In this section, we comprehensively evaluate OmniForcing against both bidirectional models and cascaded autoregressive baselines. Since our primary goal is to achieve high-fidelity streaming outputs, we focus our evaluation on visual generation quality, audio fidelity, and real-time inference efficiency.
\subsection{Experimental Setup}

\noindent\textbf{Implementation Details.}
We build OmniForcing on top of LTX-2~\cite{hacohen2026ltx} (14B video stream + 5B audio stream). Training uses 32 GPUs with bf16 precision, a global batch size of 32, and a learning rate of $2 \times 10^{-5}$. Stage~I (Bidirectional DMD) and Stage~III (Joint Self-Forcing DMD) each run for 2{,}000 steps; Stage~II (Causal ODE Regression) runs for 3{,}000 steps. DMD training in Stages~I and III employs backward simulation~\cite{yin2024improved}. We use $S\!=\!16$ Audio Sink Tokens with Identity RoPE from Stage~II onward and set classifier-free guidance scales to $w_v\!=\!3$, $w_a\!=\!5$. Our training set combines Mixkit video clips with captions sourced from the Open-Sora-Plan project~\cite{lin2024open}, along with audio captions from AudioCaps~\cite{kim2019audiocaps}. All captions are rewritten by Gemma~3~12B~\cite{kamath2025gemma} to produce audio-visually coherent descriptions. Since our method distills from a pretrained teacher rather than training from scratch, this relatively compact dataset suffices for the distillation objectives.

\noindent\textbf{Evaluation Metrics.}
We evaluate on JavisBench~\cite{liu2026javisdit}, following its four-dimension protocol: \textit{AV-Quality} (FVD~\cite{unterthiner2018towards}, FAD~\cite{kilgour2018fr}), \textit{Text-Consistency} (TV-IB, TA-IB via ImageBind~\cite{girdhar2023imagebind}; CLIP~\cite{radford2021learning}; CLAP~\cite{elizalde2023clap}), \textit{AV-Consistency} (AV-IB, AVHScore), and \textit{AV-Synchrony} (JavisScore~\cite{liu2026javisdit}, DeSync~\cite{iashin2024synchformer}). We additionally report Runtime, TTFC, and FPS to quantify streaming capability. Here, TTFC measures the wall-clock time to generate and decode the Global Prefix ($\mathcal{B}_0$) together with the first streaming block ($\mathcal{B}_1$); once these are emitted, subsequent blocks are decoded concurrently with generation, enabling uninterrupted streaming playback. For distillation fidelity analysis, we use VBench~\cite{huang2024vbench}, which officially supports per-video evaluation on user-supplied content.

\noindent\textbf{Baselines.}
We compare against methods across three paradigms (\cref{tab:main_results}): \textit{T2A+A2V} cascaded pipelines (TempoTokens~\cite{yariv2024diverse}, TPoS~\cite{jeong2023power}), \textit{T2V+V2A} cascaded pipelines (ReWaS~\cite{jeong2025read}, Seeing\&Hearing~\cite{xing2024seeing}, FoleyCrafter~\cite{zhang2026foleycrafter}, MMAudio~\cite{cheng2025mmaudio}), and \textit{T2AV} joint models (MM-Diffusion~\cite{ruan2023mm}, JavisDiT~\cite{liu2025javisdit}, UniVerse-1~\cite{wang2025universe}, JavisDiT++~\cite{liu2026javisdit}). The bidirectional LTX-2~\cite{hacohen2026ltx} serves as the teacher upper bound.

\begin{figure}[!t]
\centering
\includegraphics[width=\textwidth]{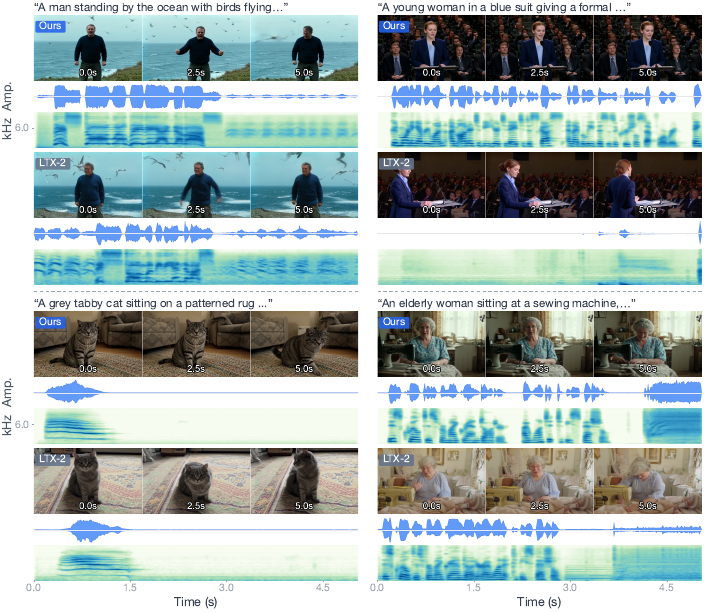}
\caption{Qualitative comparison across diverse scenes. Each example 
shows generated frames with synchronized waveforms and Mel spectrograms. 
OmniForcing produces voice layered with bird calls at a seaside scene 
(top-left), sustained speech at a podium presentation (top-right), a 
precisely timed cat meow (bottom-left), and blended narration with 
sewing machine sounds (bottom-right).}
\label{fig:demo_showcase}
\end{figure}
\subsection{Main Results}

\noindent\textbf{Inference Efficiency.}
As shown in Tables~\ref{tab:main_results} and~\ref{tab:distill_fidelity}, OmniForcing completes a 5-second 480p audio-visual clip in ${\sim}5.7$\,s of wall-clock time, with a TTFC of ${\sim}0.7$\,s and a sustained throughput of ${\sim}25$\,FPS---a ${\sim}35{\times}$ speedup over the offline LTX-2 teacher (197\,s) and the only method in our comparison that enables true streaming generation. 

\noindent\textbf{Audio-Visual Quality.}
On JavisBench (\cref{tab:main_results}), OmniForcing attains an FVD of 137.2 and FAD of 5.7, surpassing all baselines except the bidirectional teacher (FVD 125.4, FAD 4.6) on FVD and remaining comparable to the strongest competitor JavisDiT++ (FVD 141.5, FAD 5.5) on FAD, demonstrating that the core distributional quality of both video and audio is well preserved through distillation. For text-consistency, OmniForcing achieves the best CLIP score (0.322) across all methods, surpassing both LTX-2 (0.318) and JavisDiT++ (0.316), while TV-IB (0.287) ranks second overall. The audio-text metrics TA-IB (0.162) and CLAP (0.401) are slightly below the teacher but remain competitive with JavisDiT++ (0.164 and 0.424). For cross-modal coherence, OmniForcing achieves an AV-IB of 0.269 and AVHScore of 0.254, both ranking second overall and substantially outperforming all non-teacher baselines (the nearest competitor scores 0.198 on AV-IB). Temporal synchronization is also well preserved, with a DeSync of 0.392 closely tracking the teacher's 0.384 and dramatically outperforming all cascaded and joint baselines (JavisDiT++: 0.832). We attribute the modest gaps relative to the teacher in consistency and synchrony to the restricted causal receptive field replacing the original bidirectional full-sequence attention, an inherent trade-off for achieving streaming capability. Overall, OmniForcing achieves a ${\sim}35{\times}$ runtime reduction while attaining the strongest or second-strongest scores across nearly all quality dimensions.

\noindent\textbf{Further Results on VBench and Qualitative Analysis.}
Since comprehensive joint audio-visual benchmarks remain scarce, we further validate the visual fidelity of our distillation using VBench~\cite{huang2024vbench} (\cref{tab:distill_fidelity}). Under a controlled same-prompt comparison, OmniForcing slightly outperforms the teacher on per-frame quality metrics, including aesthetic quality ($+0.026$), imaging quality ($+0.020$), and subject consistency ($+0.010$), while temporal coherence (motion smoothness, temporal flickering) is likewise preserved. Although this may appear counterintuitive, it is consistent with prior distillation work~\cite{yin2024improved,huang2025self}, which report that DMD-distilled students can surpass their teachers on per-sample quality metrics. On the audio side, qualitative results in \cref{fig:demo_showcase} show that OmniForcing synthesizes layered voice-and-ambient sounds, sustained speech, synchronized transient events, and complex audio blends comparable to the teacher. These results together confirm that OmniForcing achieves a ${\sim}35{\times}$ latency reduction while maintaining the teacher's perceptual generation fidelity.

\begin{table}[t]
\centering
\caption{Main results on JavisBench~\cite{liu2025javisdit}. Best results in \textbf{bold}, second-best \underline{underlined}. ($\uparrow$: higher is better; $\downarrow$: lower is better). TTFC and FPS metrics are reported in \cref{tab:distill_fidelity}.}
\label{tab:main_results}
\resizebox{\textwidth}{!}{
\begin{tabular}{lcccccccccccc}
\toprule
& & \multicolumn{2}{c}{AV-Quality} & \multicolumn{4}{c}{Text-Consistency} & \multicolumn{2}{c}{AV-Consistency} & \multicolumn{2}{c}{AV-Synchrony} & \\
\cmidrule(lr){3-4} \cmidrule(lr){5-8} \cmidrule(lr){9-10} \cmidrule(lr){11-12}
Model & Size & FVD $\downarrow$ & FAD $\downarrow$ & TV-IB $\uparrow$ & TA-IB $\uparrow$ & CLIP $\uparrow$ & CLAP $\uparrow$ & AV-IB $\uparrow$ & AVHScore $\uparrow$ & JavisScore $\uparrow$ & DeSync $\downarrow$ & Runtime $\downarrow$ \\
\midrule
\textit{-- T2A+A2V} & & & & & & & & & & & & \\
TempoTkn & 1.3B & 539.8 & -- & 0.084 & -- & 0.205 & -- & 0.139 & 0.122 & 0.103 & 1.532 & 20s \\
TPoS & 1.0B & 839.7 & -- & 0.201 & -- & 0.229 & -- & 0.124 & 0.129 & 0.095 & 1.493 & 19s \\
\midrule
\textit{-- T2V+V2A} & & & & & & & & & & & & \\
ReWaS & 0.6B & -- & 9.4 & -- & 0.123 & -- & 0.280 & 0.110 & 0.104 & 0.079 & 1.071 & 17s \\
See\&Hear & 0.4B & -- & 7.6 & -- & 0.129 & -- & 0.263 & 0.160 & 0.143 & 0.112 & 1.099 & 25s \\
FoleyC & 1.2B & -- & 9.1 & -- & 0.149 & -- & 0.383 & 0.193 & {0.186} & 0.151 & 0.952 & 16s \\
MMAudio & 0.1B & -- & {6.1} & -- & {0.160} & -- & {0.407} & {0.198} & 0.182 & 0.150 & {0.849} & 15s \\
\midrule
\textit{-- T2AV} & & & & & & & & & & & & \\
MM-Diff & 0.4B & 2311.9 & 27.5 & 0.080 & 0.014 & 0.181 & 0.079 & 0.119 & 0.109 & 0.070 & 0.875 & \underline{9s} \\
JavisDiT & 3.1B & 204.1 & 7.2 & 0.263 & 0.143 & 0.302 & 0.391 & 0.197 & 0.179 & 0.154 & 1.039 & 30s \\
UniVerse-1 & 6.4B & {194.2} & 8.7 & {0.272} & 0.111 & {0.309} & 0.245 & 0.104 & 0.098 & 0.077 & 0.929 & 13s \\
JavisDiT++ & 2.1B & {141.5} & \underline{5.5} & {0.282} & \underline{0.164} & {0.316} & \underline{0.424} & {0.198} & 0.184 & {0.159} & 0.832 & 10s \\
\midrule
\textit{-- T2AV \& Autoregressive} & & & & & & & & & & & & \\
LTX-2 & 19B & \textbf{125.4} & \textbf{4.6} & \textbf{0.290} & \textbf{0.173} & \underline{0.318} & \textbf{0.442} & \textbf{0.318} & \textbf{0.298} & \textbf{0.253} & \textbf{0.384} & 197s \\
\textbf{Ours} & 19B & \underline{137.2} & 5.7 & \underline{0.287} & 0.162 & \textbf{0.322} & 0.401 & \underline{0.269} & \underline{0.254} & \underline{0.208} & \underline{0.392} & \textbf{5.7s} \\
\bottomrule
\end{tabular}
}
\end{table}

\begin{table}[t]
\centering
\caption{Distillation fidelity: OmniForcing vs.\ the bidirectional LTX-2 teacher on the same prompt set. Visual quality is measured using VBench~\cite{huang2024vbench}, which provides a modular toolkit that officially supports per-video scoring on user-supplied content. Best in \textbf{bold}. ($\uparrow$: higher is better; $\downarrow$: lower is better).}
\label{tab:distill_fidelity}
\resizebox{\textwidth}{!}{
\begin{tabular}{lcccccccc}
\toprule
& \multicolumn{5}{c}{VBench Visual Quality} & \multicolumn{2}{c}{Streaming} \\
\cmidrule(lr){2-6} \cmidrule(lr){7-8}
Model & Aesthetic Quality $\uparrow$ & Imaging Quality $\uparrow$ & Motion Smoothness $\uparrow$ & Subject Consistency $\uparrow$ & Temporal Flickering $\uparrow$ & TTFC $\downarrow$ & FPS $\uparrow$ \\
\midrule
LTX-2~\cite{hacohen2026ltx} & 0.569 & 0.574 & 0.993 & 0.945 & 0.988 & 197.0s & -- \\
\textbf{OmniForcing} & \textbf{0.595} & \textbf{0.594} & \textbf{0.995} & \textbf{0.955} & \textbf{0.989} & \textbf{0.7s} & \textbf{25} \\
\bottomrule
\end{tabular}
}
\end{table}
\subsection{Ablation Studies}

We isolate key components during Stage~II, where the model first encounters the causal mask and instability is most pronounced. Results are summarized in \cref{tab:ablation}.

\noindent\textbf{Audio Attention Sink and Identity RoPE.}
We sweep the sink pool size $S \in \{1,2,4,8,16,24\}$. Configurations with $S \ge 4$ converge stably with normal visual quality, while $S \le 2$ triggers NaN gradients from Softmax collapse (Table~\ref{tab:ablation}), confirming that a sufficiently large sink pool is necessary to restore attention entropy. Replacing Identity RoPE with standard incremental positions at $S\!=\!16$ yields convergence but elevated loss (0.402 vs.\ 0.081) and noisier outputs, validating that position-agnostic anchoring is essential.

\noindent\textbf{Comparison with Alternative Stabilizers.}
QK-Norm~\cite{dehghani2023scaling} converges stably but its aggressive normalization dampens attention contrast, yielding higher loss (0.232). Tanh-Gated Attention~\cite{alayrac2022flamingo} is more problematic: although no NaN occurs, the $\tanh$ saturation suppresses gradients so heavily that the loss plateaus at 1.258 and outputs degenerate into block artifacts---a signature of functionally destroyed attention. Neither alternative matches the stability--quality trade-off of our Audio Sink Tokens.


\begin{table}[t]
\centering
\caption{Ablation on training stabilization strategies during Stage~II. We report convergence status (with failure mode if applicable), maximum gradient norm, visual quality, and one-step denoising loss at $\sigma\!=\!0.5$ averaged over the evaluation set after 3k steps. All sink variants use Identity RoPE unless noted.}
\label{tab:ablation}
\resizebox{\textwidth}{!}{
\begin{tabular}{lccccc}
\toprule
\textbf{Configuration} & \textbf{Convergence} & \textbf{$\|\nabla\|_{\max}$} & \textbf{Visualization} & \textbf{Loss ($\sigma\!=\!0.5$, 3k)} & \textbf{Remark} \\
\midrule
Sink $S\!=\!24$ + Id.\ RoPE & Stable & 9.15 & Normal & 0.110 & -- \\
Sink $S\!=\!16$ + Id.\ RoPE & Stable & 9.23 & Normal & 0.081 & -- \\
Sink $S\!=\!8$ + Id.\ RoPE  & Stable & 21.95 & Normal & 0.129 & -- \\
Sink $S\!=\!4$ + Id.\ RoPE  & Stable & 49.71 & Normal & 0.141 & -- \\
Sink $S\!=\!2$ + Id.\ RoPE  & NaN & $\infty$ & Noise & -- & Softmax collapse \\
Sink $S\!=\!1$ + Id.\ RoPE  & NaN & $\infty$ & Noise & -- & Softmax collapse \\
\midrule
Sink $S\!=\!16$ + Incr.\ RoPE & Stable & 11.21 & Noisy & 0.402 & Positional bias \\
\midrule
QK-Norm~\cite{dehghani2023scaling} & Stable & 4.45 & Normal & 0.232 & Damped contrast \\
Tanh-Gated Attn~\cite{alayrac2022flamingo} & Plateau$^\dag$ & 10.61 & Block artifacts & 1.258 & Attention destroyed \\
No Stabilizer & NaN & $\infty$ & Noise & -- & Softmax collapse \\
\bottomrule
\multicolumn{6}{l}{\small $^\dag$No NaN observed, but loss does not decrease; the $\tanh$ gate functionally destroys attention patterns.}
\end{tabular}
}
\end{table}


\section{Conclusion}
We presented OmniForcing, the first framework to distill a bidirectional joint audio-visual diffusion model into a real-time streaming autoregressive generator. To overcome the temporal asymmetry and gradient instability inherent in causal multi-modal distillation, we introduced Asymmetric Block-Causal Alignment and Audio Sink Tokens with Identity RoPE, coupled with Joint Self-Forcing Distillation and a modality-independent rolling KV-cache. OmniForcing achieves $\sim$25 FPS streaming on a single GPU. We hope this work opens up new possibilities for deploying multi-modal foundation models in interactive and latency-sensitive scenarios.



%
%
\bibliographystyle{splncs04}
\bibliography{main}
\end{document}